\documentclass[aps,pra,twocolumn,superscriptaddress]{revtex4-1}
\usepackage[utf8]{inputenc}
\usepackage{graphicx}
\usepackage[colorlinks, citecolor={blue}, urlcolor={blue}, linkcolor={blue}]{hyperref}
\usepackage{amsmath}
\usepackage[dvipsnames]{xcolor}
\newcommand{\spindown}{|\hspace{-2pt}\downarrow\rangle}
\newcommand{\spinup}{|\hspace{-2pt}\uparrow\rangle}

\newcommand{\ket}[1]{\left | #1 \right \rangle}

\newcommand{\UMD}{Joint Quantum Institute and Department of Physics, University of Maryland, College Park, MD 20742}
\newcommand{\DQC}{Duke Quantum Center, Department of Electrical and Computer Engineering and Department of Physics, Duke University, Durham, NC 27708}
\newcommand{\IonQ}{IonQ, Inc., College Park, MD 20740}

\begin{document}

\title{Engineering magnetically insensitive qubits \\ in metastable electronic D-states of trapped ions}

\author{Ksenia Sosnova}
\email{Corresponding Author: ksosnova@ionq.co}
\affiliation{\IonQ}
\affiliation{\UMD}

\author{Martin Lichtman}
\email{Present Address: Infleqtion, Inc., Madison, WI 53703}
\affiliation{\UMD}

\author{Allison Carter}
\email{Present Address: National Institute of Standards and Technology, Boulder, CO  80305}
\affiliation{\UMD}

\author{Nora Crocker}
\email{Present Address: Keysight Technologies, Santa Rosa, CA 95403}
\affiliation{\UMD}

\author{Christopher Monroe}
\affiliation{\IonQ}
\affiliation{\UMD}
\affiliation{\DQC}

\date{\today}

\begin{abstract}
Ion trap quantum computers often store qubits on field-sensitive $S_{1/2}$ ground state Zeeman levels of the valence electron, such as in $^{40}$Ca$^+$, $^{88}$Sr$^+$, and $^{138}$Ba$^+$ atomic systems. 
We experimentally synthesize magnetically insensitive qubit states in multiple metastable electronic $D_{3/2}$ Zeeman levels in such an atomic system. We demonstrate coherent operations within the $D_{3/2}$ manifold of $^{138}$Ba$^+$, including coherent flopping between the synthesized qubit states, and our results agree with theory. Such an encoding may allow for more flexible use of atomic levels for photonic interfaces, and with a measured improvement in the qubit coherence time $T_2^*$ by a factor of $3$, this lays the foundation for further improvement for quantum computing and network applications. 
\end{abstract}

\maketitle

\section{Introduction}
A leading platform for scalable quantum computing \cite{Alexeev2021} and quantum communication networks~\cite{Awschalom2021} is a collection of trapped atomic ions, with their unsurpassed coherence times and high-fidelity operations. 
Hyperfine $S_{1/2}$ ground ``clock" states are popular choices for qubit levels, with their low sensitivity to magnetic field fluctuations \cite{Wang2021}. Some atomic ion species have no nuclear spin and hence lack hyperfine structure and clock states. However, combinations of excited metastable levels in these atoms can offer effective magnetic insensitivity. Such metastable levels also allow for the storage of an additional type of qubit in the same atom, to separate idle quantum memory from dissipative operations such as sympathetic cooling or photonic interfaces \cite{allcockOmgAppliedPhysicsLetters2021, Yu2025, Toh2025}. Finally, excited metastable qubit levels usually come with longer-wavelength optical transitions for compatibility with a wide array of integrated photonic devices not available in the blue or ultraviolet spectrum.

\begin{figure}
 \includegraphics[width=\linewidth]{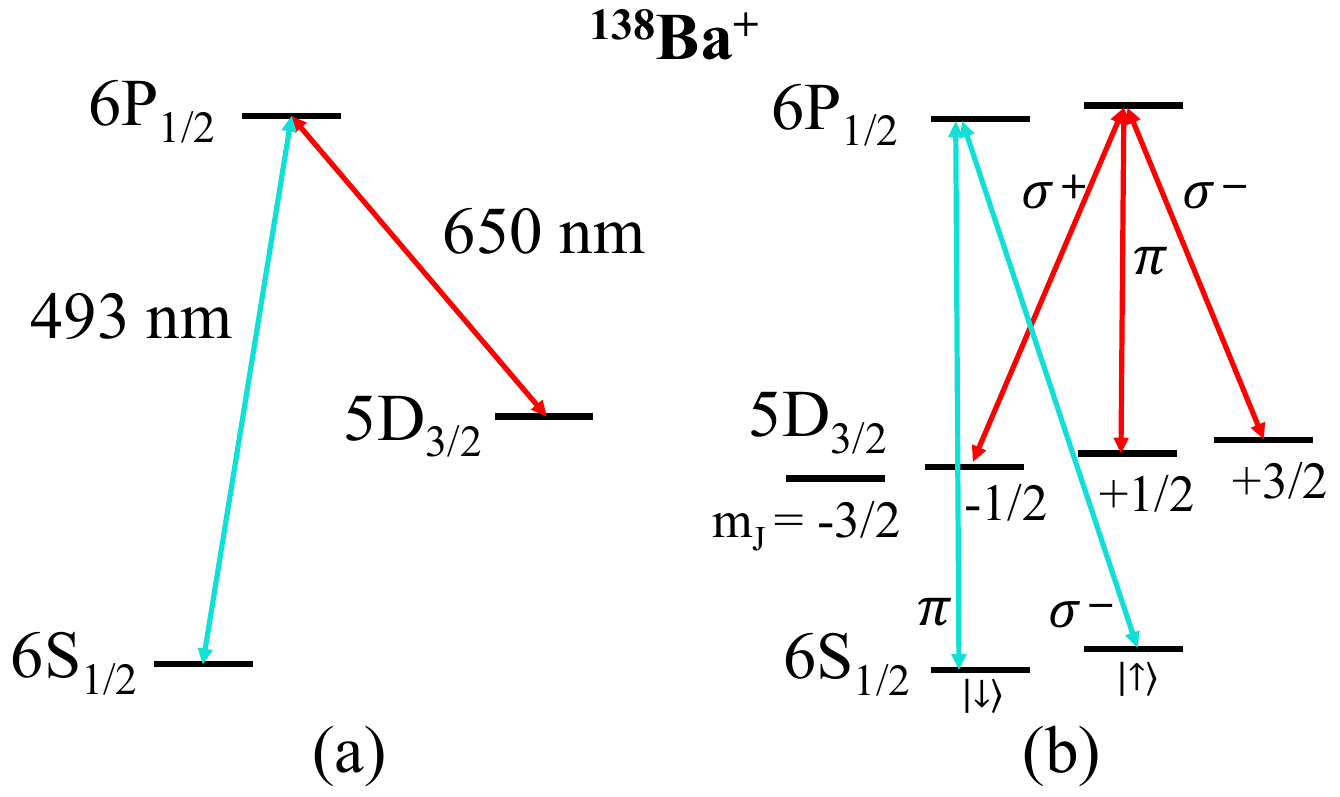}
 \caption{\label{fig_scheme} $^{138}$Ba$^+$ low-lying electronic structure (a) in the absence of magnetic field, and (b) in the presence of weak magnetic field, splitting the levels by their projection of angular momentum along the field given by quantum number $m_J$. A few representative transitions are shown for clarity.}
\end{figure}

Here we consider the nuclear spinless $^{138}$Ba$^+$ atomic ion, with its lowest energy states indicated in Fig. \ref{fig_scheme}. The $S_{1/2}$ ground state is a Zeeman doublet with frequency separation $2\mu_B B$ MHz/G, where $\mu_B = 1.4$ MHz/G is the electron Bohr magneton and $B$ is the magnetic field. The excited $P_{1/2}$ state, with lifetime $7.86$ ns, allows for fast optical pumping (initialization) and fluorescence (detection) operations.
The metastable $D_{3/2}$ state has a natural lifetime of $80$ s \cite{Yu1997}, which is much longer than any quantum operation considered here. In addition, the branching ratio from the $P_{1/2}$ manifold to the $S_{1/2}$ and $D_{3/2}$ manifolds is 3:1 (see Fig.~\ref{fig_scheme}a), ensuring fast rates of pumping into and exciting out of the $D_{3/2}$ states. Moreover, the 650~nm $P_{1/2} - D_{3/2}$ transition is spectrally distant from the 493~nm $S_{1/2} - P_{1/2}$ transition, ensuring no optical crosstalk, and both transition colors are compatible with fiber and integrated optical technologies.
The storage and detection of the $D_{3/2}$ states has not yet been widely used in quantum information applications, although there have been several theoretical proposals to store and manipulate qubits in such atomic states~\cite{Aharon2013, Aharon2017, Aharon2016}.

In this paper, we introduce a novel detection technique for all four states in the $D_{3/2}$ manifold of the $^{138}$Ba$^+$ ion, and this general scheme can also be applied to other ions with nuclear spin $I = 0$, such as $^{40}$Ca$^+$~\cite{Chwalla2009, Liu2011}, and $^{88}$Sr$^+$~\cite{Akerman2012, Letchumanan2007}. We also demonstrate various coherent operations in the $D_{3/2}$ manifold including coherent $\Delta m_J = \pm 1$ and $\Delta m_J = \pm 2$ rotations within the manifold, showing good agreement with theoretical simulations. 

With a pair of Raman beams or, alternatively, a STIRAP procedure using 650~nm light, we synthesize a qubit with reduced magnetic field sensitivity, and show improvement in the coherence time $T^*_2$ by the factor of 3 in comparison to a magnetically sensitive qubit. We discuss future steps in improving the coherence time by creating a protected subspace and applying a driving Hamiltonian that constantly reprojects the ion state into this subspace~\cite{Aharon2013, Aharon2017, Aharon2016}.

\section{Detection} \label{detection}

The $^{138}$Ba$^+$ qubit is typically defined in the $6S_{1/2}$ ground state manifold with the electron spin states $\spindown$ $(m_J=-1/2)$ and $\spinup$ $(m_J=+1/2)$, where $m_J$ is the projection of the electron angular momentum along the (magnetic field) quantization axis. Recently, we demonstrated a probabilistic detection scheme for these qubit states~\cite{Inlek2017}. For this detection procedure, we turn on all polarizations of 650~nm light (driving all $\Delta m_J = 0$ and $\pm 1$ transtions) and either the $\sigma^{+}$ or $\sigma^{-}$ polarization of 493~nm light (driving $\Delta m_J = \pm 1$ transitions, respectively). We then repeat the experiment using the opposite polarization of 493~nm light. If, for example, the ion is in state $\spinup$ and exposed to only 650~nm light and 493~nm $\sigma^{+}$ light, the ion will not scatter any photons. If, however, the ion is in state $\spindown$, the application of the same laser beams will scatter, on average, $2.8$ 493~nm photons from the ion before it is pumped into the dark $\spinup$ state. By counting the number of scattered 493~nm photons over the course of many experimental repetitions, we are thus able to determine the populations of the two $S_{1/2}$ manifold states (see Appendix~\ref{append_S} for more details). This scheme exhibits a single-shot qubit detection efficiency of only $8\%$, but this is not necessarily a problem in some quantum information protocols.  For example, communication qubits need not be measured in the ultimate implementation of an ion trap quantum network.  An alternative detection scheme shelves one of the ground state qubit states to  another long-lived metastable state such as $5D_{5/2}$~\cite{Dietrich2010, Keselman2011, OReilly2024}.

Here, we extend this detection scheme to the populations in each of the four Zeeman levels in the $5D_{3/2}$ manifold. Instead of using the 650~nm light as a repumping beam and alternating 493~nm polarizations, we now leave all polarizations of the 493~nm light on during detection to repump from the $6S_{1/2}$ manifold and cycle through different 650~nm polarization settings. While the two Zeeman levels in the $6S_{1/2}$ manifold required only two different 493~nm polarizations to determine their populations, the greater complexity of the $5D_{3/2}$ state requires additional polarization combinations. We use a succession of five linearly independent polarization settings to determine the population of the four Zeeman levels: (i) $\sigma^{+}$, (ii) $\sigma^{-}$, (iii) $\pi$, (iv) combination $\sigma^{+}$ and $\pi$, and (v) combination $\sigma^{-}$ and $\pi$. Intuitively, we can understand the linear independence of each polarization or polarization combinations by considering the dark states of the ion when each polarization is applied. For $D_{3/2} \xrightarrow{} P_{1/2}$ ($J=3/2\xrightarrow{}J=1/2$) transitions, there are two dark states for any polarization of the applied light~\cite{Berkeland2002}. For an individual polarization, $\sigma^{+}$, $\sigma^{-}$, or $\pi$, both dark states are stationary even when a magnetic field is applied. However, for the combinations $\sigma^{\pm}$ and $\pi$, only one dark state is stationary, while the other quickly evolves to a bright state if the $\pi$ and $\sigma^{\pm}$ beams have different detunings from the transition resonance.

Each time we run an experiment involving detection in the $5D_{3/2}$ manifold, we prepare the same state 5 times and measure the 493~nm  fluorescence given each of the above 5 polarization settings. By comparing the number of average photons collected for each polarization combination after many trials, we are able to calculate the populations in each of the $5D_{3/2}$ Zeeman levels. We constrain our results such that the sum of the four $D_{3/2}$ populations $d_{m_J}$ of the levels add to 1, with $0 \le d_{m_J} \le 1$.  Details on solving the resulting over-constrained matrix problem are found in Appendix~\ref{append_D}. In the remainder of this paper, we utilize this novel detection protocol to demonstrate rotations in the $5D_{3/2}$ manifold and evaluate a magnetically-insensitive qubit comprised of superpositions of $5D_{3/2}$ manifold Zeeman levels.

\section{Rotations}\label{sec:rotations}

 \begin{figure}
 \includegraphics[width=0.7\linewidth]{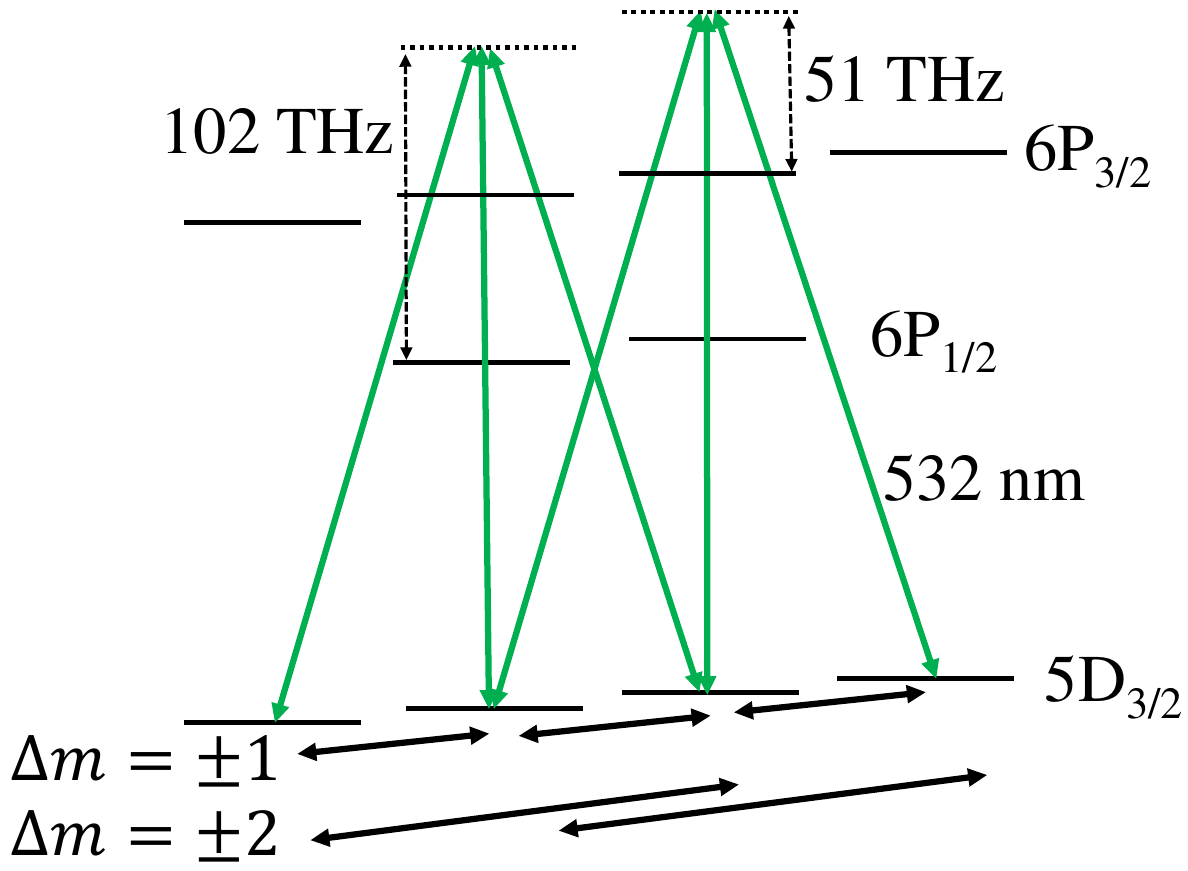}
 \caption{\label{fig_rotations} Stimulated Raman coupling of 532~nm laser beams between $D_{3/2}$ states -- represented by the green arrows -- through $P_{1/2}$ and $P_{3/2}$ manifolds in $^{138}$Ba$^{+}$. Short black arrows represent $\Delta m_J = \pm 1$ rotations between all four $D_{3/2}$ states, requiring $\sigma^\pm$ and $\pi$ polarizations of the 532~nm light. Long black arrows represent $\Delta m_J = \pm 2$ rotations between the two possible pairs of states, requiring $\sigma^+$ and $\sigma^-$ polarizations. In order to minimize the two-photon AC Stark shifts, we keep the $\pi$ polarization component on for these operations as well.}
 \end{figure}
 
In the $D_{3/2}$ manifold, we can perform two types of qubit rotations: (i) $\Delta m_J = \pm 1$, and (ii) $\Delta m_J = \pm 2$, as shown in Fig.~\ref{fig_rotations}. 
The splitting for the $\Delta m_J = \pm 1$ transitions is given by the Zeeman splitting between adjacent levels, $4\mu_B B/5$, with $B$ is the magnetic field in Gauss. Similarly, the frequency for $\Delta m_J = \pm 2$ transitions is $8\mu_B B/5$. For all the experiments discussed in this manuscript, we used the magnetic field $B = 2.2$~G, so the frequency between the adjacent levels is approximately $2.5$~MHz.

Coherent operations with $\Delta m_J = \pm 1$ can be implemented by applying RF directly to the trap electrodes \cite{Schafer2018}, or by using an inductive coil near the trap \cite{Sherman2013}. In order to perform both $\Delta m_J = \pm 1$ and $\Delta m_J = \pm 2$ rotations, we utilize a pair of 532~nm Raman co-propagating beams that are off-resonantly coupled to the $6P_{1/2}$ and $6P_{3/2}$ states.

 \begin{figure}
 \includegraphics[width=\linewidth]{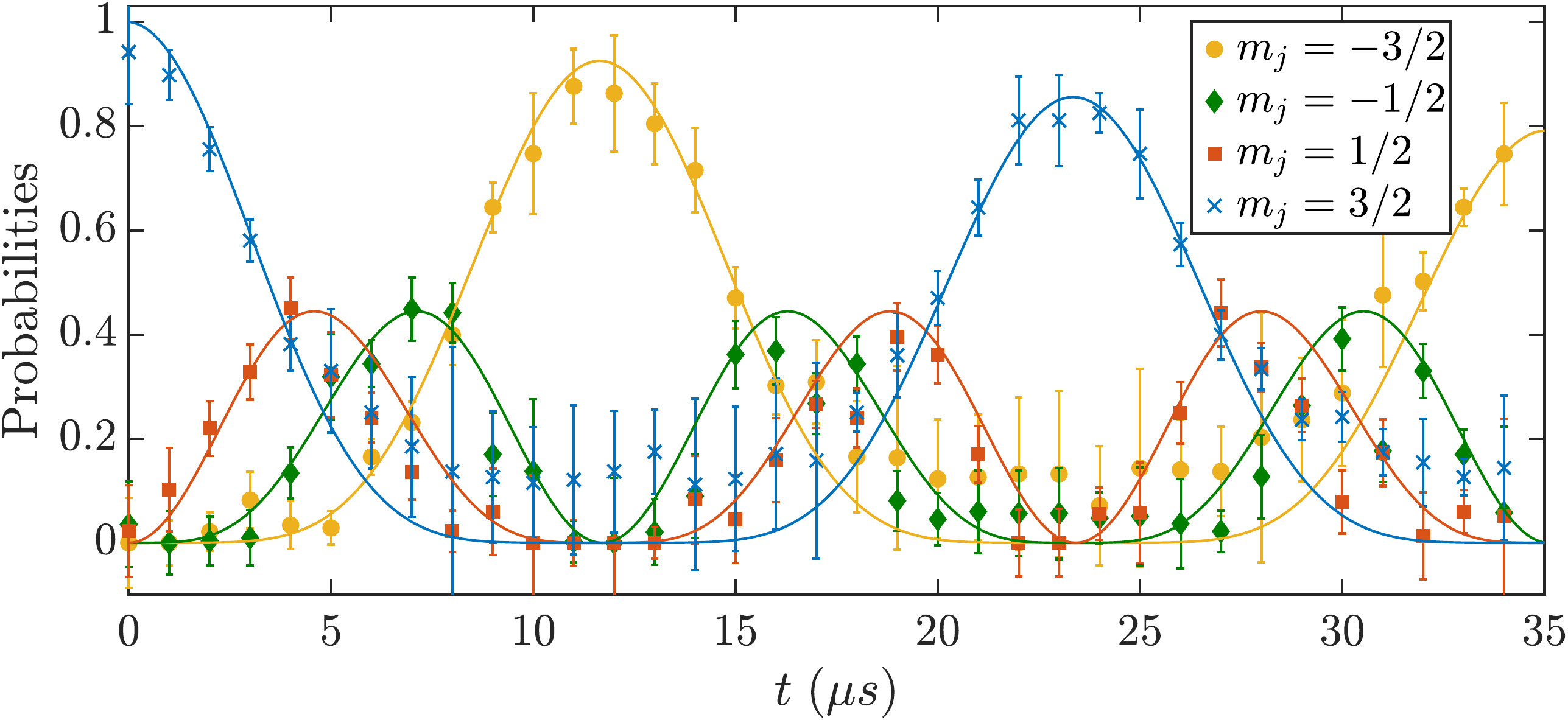}
 \caption{\label{fig_45} State populations during coherent rotations with $\Delta m_J = \pm 1$ in the $D_{3/2}$ manifold. In the presence of the pair of 532~nm Raman beams with all polarizations, all four states are involved in the time evolution. The markers show the experimental data. Yellow circles: $\ket{d_{-3/2}}$; green diamonds: $\ket{d_{-1/2}}$; red squares: $\ket{d_{+1/2}}$; and blue crosses: $\ket{d_{+3/2}}$. Solid lines of the corresponding colors show theoretical simulations for a four-level system initialized with all the population in the $\ket{d_{+3/2}}$ state with the effective microwave rotations between the $D_{3/2}$ states. We fit the theory to the experimental data over the Rabi frequency $\Omega$ and the decay time $\tau$.}
 \end{figure}

We start with $\Delta m_J = \pm 1$ rotations in the $D_{3/2}$ manifold shown in Fig.~\ref{fig_45}. \textcolor{black}{In order to minimize differential two-photon AC Stark shifts, we adjust the polarizations of our 532~nm Raman beams so that the intensities of each polarization ($\sigma^+$, $\sigma^-$, and $\pi$) are all equal}. We initialize the qubit in the $\ket{d_{+3/2}}$ state, and by applying two 532~nm Raman beams, we drive the populations between all Zeeman $D_{3/2}$ states. The experimental data agrees with the theoretical evolution simulations within the uncertainty.

The theoretical simulations use the QuTiP (Quantum Toolbox in Python) package and are performed for a four-level system with the effective $\Delta m_J = \pm 1$ microwave rotations that are described by the following interaction Hamiltonian:
\begin{equation}
    H_{\Delta m = \pm 1} = \begin{pmatrix} 
   0 & \sqrt{\dfrac{3}{18}}\dfrac{\Omega}{2} & 0 & 0  \\
   \sqrt{\dfrac{3}{18}}\dfrac{\Omega}{2} & 0 & \sqrt{\dfrac{4}{18}}\dfrac{\Omega}{2} & 0  \\
   0 & \sqrt{\dfrac{4}{18}}\dfrac{\Omega}{2} & 0 & \sqrt{\dfrac{3}{18}}\dfrac{\Omega}{2} \\
   0 & 0 & \sqrt{\dfrac{3}{18}}\dfrac{\Omega}{2} & 0  \\
   \end{pmatrix} .
\end{equation}
We take into account the Clebsch-Gordan coefficients and all possible pairs of Raman beam polarizations (see Fig.~\ref{fig_rotations}) when calculating the effective Rabi frequency.
In our setup, it takes 12~$\mu$s to transfer all the population from one of the edge states, $\ket{d_{+3/2}}$, to \textcolor{black}{the other} one, $\ket{d_{-3/2}}$, through the middle states.
 
\begin{figure}
 \includegraphics[width=\linewidth]{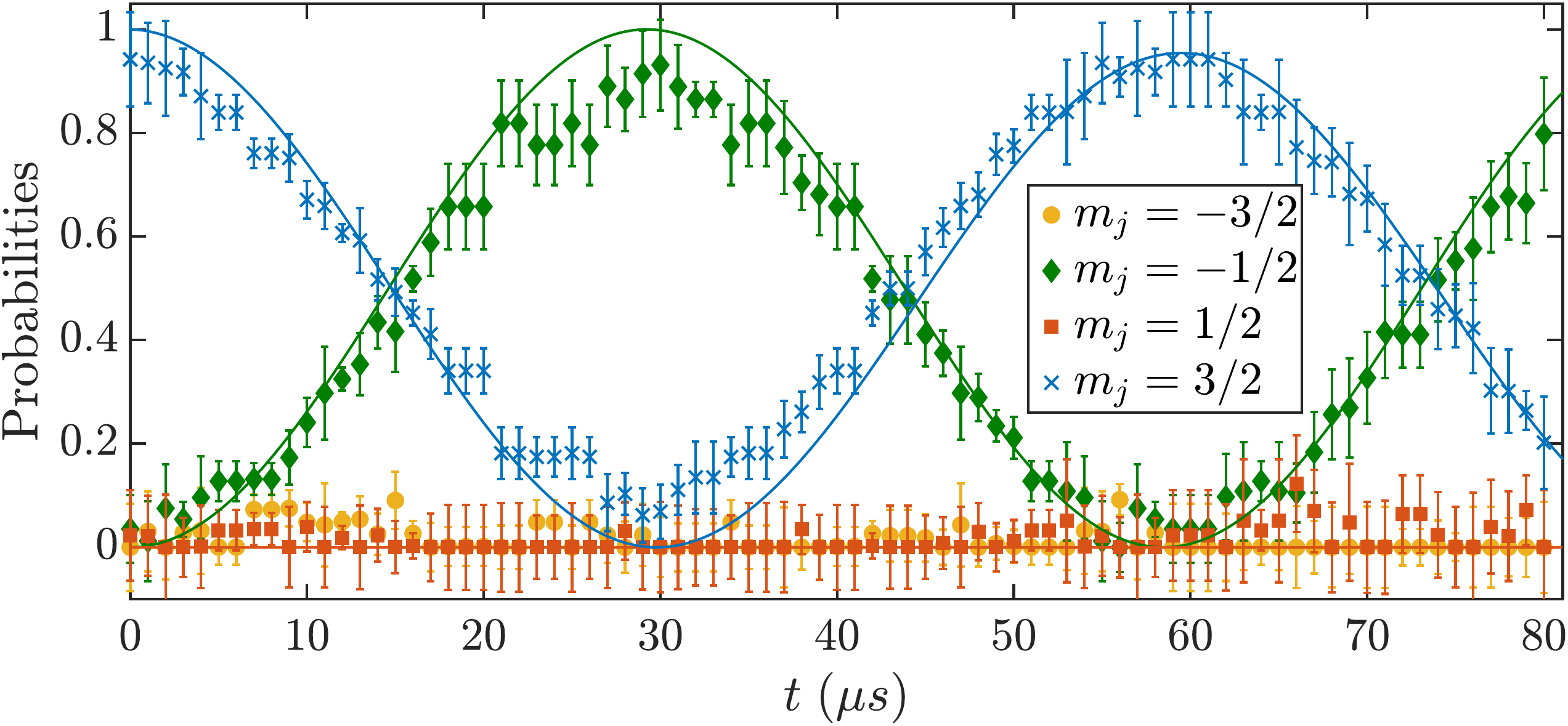}
 \caption{\label{fig_85} State populations during coherent rotations with $\Delta m_J = \pm 2$ in the $D_{3/2}$ manifold. In the presence of a pair of 532~nm Raman beams with $\sigma^+$ and $\sigma^-$ polarizations, only two states are evolving while the other two are unaffected. The markers show the experimental data. Yellow circles: $\ket{d_{-3/2}}$; green diamonds: $\ket{d_{-1/2}}$; red squares: $\ket{d_{+1/2}}$; and blue crosses: $\ket{d_{+3/2}}$. Solid lines of the corresponding colors show the QuTiP simulations for a four-level system initialized with all the population in the $\ket{d_{+3/2}}$ state with the effective microwave rotations between the $D_{3/2}$ states. We fit the QuTiP simulations to the experimental data over the Rabi frequency $\Omega$ and the decay time $\tau$.}
 \end{figure}

Next, in Fig.~\ref{fig_85}, we demonstrate $\Delta m_J\nobreak=\nobreak\pm 2$ rotations in the $D_{3/2}$ manifold. 532~nm Raman beams with $\sigma^+$ and $\sigma^-$ polarizations are applied. As in the previous case, we initialize the qubit in the $\ket{d_{+3/2}}$ state. \textcolor{black}{Now, however, }we tune the frequency difference between two Raman beams to $\dfrac{8}{5}\mu_B B$, and perform rotations only between two states, $\ket{d_{+3/2}}$ and $\ket{d_{-1/2}}$, while other two states are unpopulated. We additionally check the rotations between the second pair of states: $\ket{d_{-3/2}}$ and $\ket{d_{+1/2}}$. The experimental data agrees with the QuTiP simulations within the uncertainty in this case, as well. 

The QuTiP simulations are performed for a four-level system with the effective $\Delta m_J = \pm 2$ microwave rotations that are described by the following interaction Hamiltonian:
\begin{equation}
    H_{\Delta m = \pm 2} = \begin{pmatrix} 
   0 & 0 & \dfrac{\Omega}{2} & 0  \\
   0 & 0 & 0 & \dfrac{\Omega}{2}  \\
   \dfrac{\Omega}{2} & 0 & 0 & 0 \\
   0 & \dfrac{\Omega}{2} & 0 & 0  \\
   \end{pmatrix} .
\end{equation}
 As shown in Fig.~\ref{fig_85}, we can perform a $\pi/2$ rotation in $T = 30~\mu$s, and as a result, end up in the $\ket{d_{-1/2}}$ state. \textcolor{black}{Therefore,} at $t = \dfrac{2}{3}T = 20~\mu$s, the electron will be in the state: $\dfrac{1}{2}\ket{d_{+3/2}} + e^{\rm{i}\phi}\dfrac{\sqrt{3}}{2}\ket{d_{-1/2}}$, where the phase $\phi$ is defined by the phase shift between the two Raman beams. Let $|D_1\rangle$ denote such a state with $\phi = \pi$. Using the same phase shift and the other submanifold of $D_{3/2}$, we create an orthogonal state $|D_2\rangle$, such that:
 \begin{eqnarray}
     |D_1\rangle &=& \dfrac{1}{2}\ket{d_{+3/2}} - \dfrac{\sqrt{3}}{2}\ket{d_{-1/2}}\;,\\ \nonumber
     |D_2\rangle &=& \dfrac{1}{2}\ket{d_{-3/2}} - \dfrac{\sqrt{3}}{2}\ket{d_{+1/2}}\;.
 \end{eqnarray}
 
The above description applies to creating $|D_1\rangle$ and $|D_2\rangle$ states using a pair of Raman beams. We also implemented an alternative technique to create those states via a STIRAP procedure. In this case, we initialize the ion in the $\ket{d_{+3/2}}$ state, and use the resonant 650~nm $\sigma^{+}$ and $\sigma^{-}$ beams in the STIRAP configuration to transfer population to $\ket{d_{-1/2}}$ with the state $\ket{p_{+1/2}}$ being an excited state with the STIRAP preparation time equal $10\mu$s.

\section{Magnetically insensitive qubit}

 In contrast to a qubit defined in the $S_{1/2}$ manifold, the qubit \textcolor{black}{consisting of} the $|D_1\rangle$ and $|D_2\rangle$ states is insensitive by construction to the magnetic field fluctuations~\cite{Aharon2013, Aharon2017}: 
 \begin{equation} \label{eq:magnet-insensitive}
   \langle D_j|J_z|D_i\rangle = 0, \hspace{10pt} \forall i,j\in\{1,2\},   
 \end{equation}
 where $\boldsymbol{J}$ is the angular momentum operator.
 
 To detect the state of the new qubit with $|0\rangle \equiv |D_1\rangle$ and $|1\rangle \equiv |D_2\rangle $, we perform detection of all four states in the $D_{3/2}$ manifold as discussed in Sec.~\ref{detection}, and then we project them to the $\left\{ |D_1\rangle, |D_2\rangle\right\}$ subspace.
 
\begin{figure}
 \includegraphics[width=\linewidth]{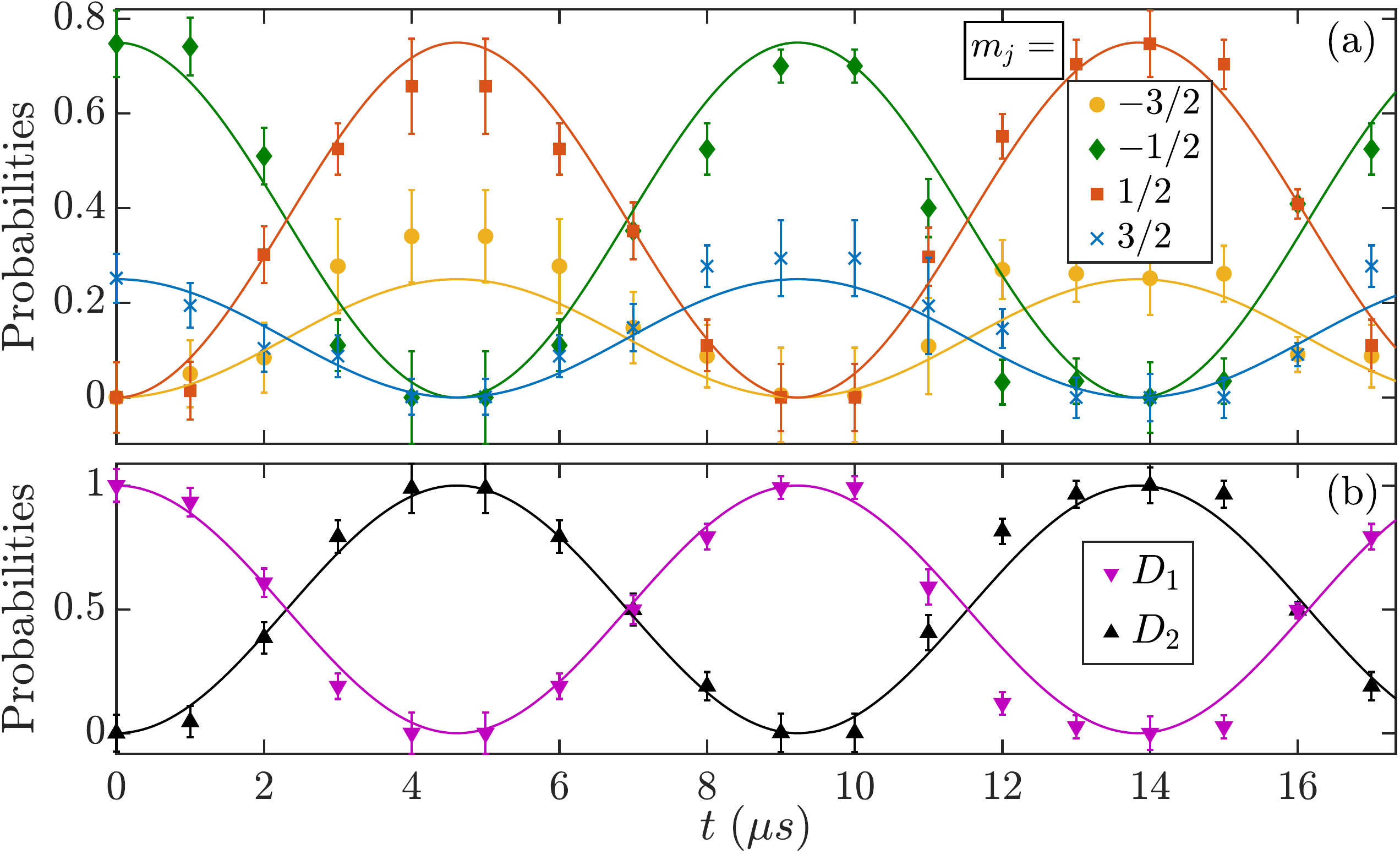}
 \caption{\label{fig_36}  Coherent $\Delta m_J = \pm 1$ rotations between the $|D_1\rangle$ and $|D_2\rangle$ states. In the presence of a pair of 532~nm Raman beams, all four $D_{3/2}$-manifold states are involved in the time evolution. {\bf (a)} The markers show the experimental data. Yellow circles: $\ket{d_{-3/2}}$; green diamonds: $\ket{d_{-1/2}}$; red squares: $\ket{d_{+1/2}}$; and blue crosses: $\ket{d_{+3/2}}$. Solid lines of the corresponding colors show the QuTiP simulations for a four-level system starting with the $|D_1\rangle$ state and performing the effective microwave rotations between the $D_{3/2}$-manifold states. We fit the QuTiP simulations to the experimental data over the Rabi frequency $\Omega$ and the decay time $\tau$. {\bf (b)} The synthetic qubit state populations. The markers show the experimental data. Magenta down-pointing triangles: $|D_1\rangle$ state; black up-ponting triangles: $|D_2\rangle$ state. Solid lines of the corresponding colors represent the corresponding QuTiP simulations.}
 \end{figure}

 The coherent $\Delta m_J = \pm 1$ rotations between $|D_1\rangle$ and $|D_2\rangle$ states can be performed by applying microwaves or by sending two off-resonantly coupled Raman beams, as discussed in the beginning of Sec.~\ref{sec:rotations}. Fig.~\ref{fig_36} shows $\Delta m_J = \pm 1$ rotations between $|D_1\rangle$ and $|D_2\rangle$ states. The experimental data agrees with the QuTiP simulations within the uncertainty.
 
 Since the new synthetic qubit is magnetically insensitive, we expect a considerable improvement in the coherence time $T^*_2$ in comparison to a conventional qubit in the $S_{1/2}$ manifold. The major limitation to the coherence time of the $S_{1/2}$-manifold qubit is magnetic noise, which mainly comes from the lab equipment and the power lines at 60~Hz and higher harmonics. There are several ways to address this issue, such as using $\mu$-metal shielding~\cite{Ruster2016} or superconducting rings~\cite{GABRIELSE1991,Wang2010}, setting up a feed-forward~\cite{Merkel2019} or feedback magnetic field stabilization~\cite{Dedman2007}, or using an alternative isotope with a magnetically insensitive hyperfine qubit, e.g. $^{133}$Ba$^+$~\cite{Hucul2017}. In our experiment, we do not perform any magnetic field stabilization or shielding. As a first benchmark, we perform a Ramsey experiment on a qubit defined in the $S_{1/2}$ manifold. In this case, $T^*_2 = (96\pm 15)~\mu$s is obtained, which is a result of high magnetic sensitivity of about $2.8$kHz/mG. As a second benchmark, we perform the same Ramsey experiment for a qubit defined in \textcolor{black}{the} $|d_3\rangle$ and $|d_1\rangle$ states from the $D_{3/2}$ manifold (the corresponding qubit rotations are shown in Fig.~\ref{fig_85}). The magnetic sensitivity for this qubit is about $2.2$kHz/mG, which \textcolor{black}{result} in a slightly longer $T^*_2 = (117\pm 11)~\mu$s. Next, we perform the Ramsey experiment on the synthetic magnetically insensitive qubit consisting of the $|D_1\rangle$ and $|D_2\rangle$ states. \textcolor{black}{Of the three Ba qubits we tested, this qubit} has the longest coherence time: $T^*_2 = (350\pm 37)~\mu$s, which is about 3 times longer than the coherence time of a magnetically sensitive qubit. The synthetic qubit shows an improvement, but these results are still far from state-of-art coherence times observed in hyperfine or optical qubits~\cite{Harty2014, Wang2021}. However, as a next step, our results can be further improved by introducing a protected qubit subspace~\cite{Aharon2013, Aharon2017}.
 When two optical 650~nm driving fields are applied, the system is continuously reprojected into the protected Hilbert subspace $\mathcal{H}_D$.
  The theoretical prediction for the coherence time of the qubit in the protected subspace is $T_2 \sim T_1 \sim 10$~s.

 In our experiment, we satisfy only the magnetic-insensitivity condition in Eq.~(\ref{eq:magnet-insensitive}). It ensures that the magnetic noise can cause transitions only between a protected state and a state in the complementary subspace, for example, the orthogonal bright states $|B_1\rangle$ and $|B_2\rangle$: 
 \begin{eqnarray}
      |B_1\rangle &=& \dfrac{1}{2}\ket{d_{-1/2}} + \dfrac{\sqrt{3}}{2}\ket{d_{+3/2}}\,,
     \\ \nonumber |B_2\rangle &=& \dfrac{\sqrt{3}}{2}\ket{d_{-3/2}} + \dfrac{1}{2}\ket{d_{+1/2}}\;.
 \end{eqnarray}
These transitions, however, can reduce the coherence time. 
As the next steps in our experiment, we will apply the resonant 650~nm driving light, satisfy the polarization, amplitude, and detuning requirements for the 650~nm driving light, and enforce phase coherence between the 650~nm driving light and the 532~nm Raman beams to establish the protected Hilbert subspace $\mathcal{H}_D$.

\section{Conclusion}
In conclusion, we introduced a novel detection method of all four states in $D_{3/2}$ manifold in a $^{138}$Ba$^+$ ion, \textcolor{black}{which we utilized to implement}  experiments \textcolor{black}{in which} we performed coherent $\Delta m_J = \pm 1$ and $\Delta m_J = \pm 2$ rotations in the $D_{3/2}$ manifold. \textcolor{black}{Additionally, we introduced experimental schemes for creating a synthetic magnetically insensitive qubit, and demonstrated a tripled} experimental coherence time $T^*_2$. \textcolor{black}{Applying the driving Hamiltonian, and thus creating a protected qubit subspace, should} further increase of the coherence time by several orders of magnitude~\cite{Aharon2013, Aharon2017}.

The synthetic protected qubit based on $^{138}$Ba$^+$ ion has a high potential for becoming an essential ingredient in the future quantum networks. Long coherence times are extremely important for quantum information processing, and the transitions in the visible and near-infrared ranges in $^{138}$Ba$^+$ are suited to take full advantage of the modern fiber and photonic technologies. Ba-based qubits can be efficiently entangled with the corresponding photonic states and can thus become the foundation for a large-scale distributed quantum-computing network~\cite{Monroe2014}, in which multiple nodes will be connected by photonic Bell-state analyzers~\cite{Maunz2007, Nolleke2013}.

\section{Acknowledgements}
We thank N. Aharon and A. Retzker for helpful discussions. This work is supported by the DOE Quantum Systems Accelerator (DE-FOA-0002253) and the NSF
STAQ Program (PHY-1818914).

\appendix

\section{Detection in $S_{1/2}$ manifold}\label{append_S}
Since there is no closed cycling transition in $^{138}$Ba$^+$ ion, we introduced a probabilistic detection method for the states in the $S_{1/2}$ manifold~\cite{Inlek2017}. Within this approach, we apply 493~nm $\sigma^+$ and $\sigma^-$ pulses and record the average number of 493~nm photons collected before the ion gets optically pumped into a dark state. For example, if the ion is in the $\spindown$ state, and we apply 493~nm $\sigma^{+}$ light only, on average the ion scatters $2.8$\; 493~nm photons before getting optically pumped into the $\spinup$ state. On the contrary, the ion in the $\spinup$ state does not scatter any photons when the $\sigma^{+}$ light is applied.

From now on, we use the notations $|s_{1/2+m_i}\rangle \equiv |S_{1/2};m_i\rangle$, where $s_i$,\; $i=0,1$, are the $S_{1/2}$ sub-level populations. These populations can be found by:
\begin{equation} \label{S-state-solution}
    s_0 = \dfrac{n_-}{n_+ + n_-}, \hspace{10pt}  s_1 = \dfrac{n_+}{n_+ + n_-},
\end{equation}
where $n_{\pm}$ represents the average number of photons collected in a $\sigma^{\pm}$ trial.
In a matrix form, it would look the following way:
\begin{equation}\label{eq:popS}
   \begin{pmatrix} 
     n_+ \\
     n_- \\
   \end{pmatrix} = E_s M^s 
   \begin{pmatrix} 
     s_0 \\
     s_1 \\
   \end{pmatrix} = E_s
   \begin{pmatrix} 
     2.8 & 0 \\
      0 & 2.8\\
   \end{pmatrix}
   \begin{pmatrix} 
     s_0 \\
     s_1 \\
   \end{pmatrix},
\end{equation}
where $E_s$ is the detection efficiency and the matrix element $M^s_{\epsilon i}$ gives the average number of 493~nm photons scattered from an ion in the state $|i\rangle$ and a trial with polarization $\epsilon$.

The solution of this matrix equation is trivial and is given in Eq.~(\ref{S-state-solution}). However, the matrix approach is useful for a similar detection method for the states in the $D_{3/2}$ manifold discussed in Appendix B.

\section{Detection in $D_{3/2}$ manifold} \label{append_D}
In this section, we generalize the probabilistic detection method to detect all four Zeeman states in the $D_{3/2}$ manifold. For that, we apply 650~nm pulses of five different polarization combinations: $\sigma^{+}$, $\sigma^{-}$, $\pi$, $\left(\sigma^+\text{ and } \pi\right)$, and $\left(\sigma^-\text{ and }\pi\right)$, while collecting 493~nm photons until the ion gets optically pumped into a dark state. During the detection procedure, the 493~nm light is turned on all the time to re-pump population out of the $S$ manifold.

We generalize Eq.~(\ref{eq:popS}) for the populations to the case of the $|d_i\rangle$ states in the $D_{3/2}$ manifold ($i=0,\dots,3 $):
\begin{equation}\label{eq:5polar}
   \begin{pmatrix} 
   n_+ \\
     n_- \\
     n_{\pi}\\
     n_{+\pi}\\
     n_{-\pi}\\
   \end{pmatrix}
  =E_d M^d 
   \begin{pmatrix} 
   d_0 \\
     d_1 \\
     d_2\\
     d_3\\
   \end{pmatrix},
\end{equation}
where $E_d$ is a detection efficiency.
$n_{\epsilon}$ represents the average number of 493~nm photons detected from a trial with the detection polarization $\epsilon\nobreak\in\nobreak\{\sigma^+, \sigma^-,\pi, (\sigma^+ \text{ and } \pi),(\sigma^- \text{ and } \pi)\}$. We perform each experiment many times and apply one of the polarization configurations $\epsilon$ to build sufficient statistics for the values of $n_{\epsilon}$. Similarly, a matrix element $M^d_{\epsilon i}$ gives the average number of 493~nm photons scattered from an ion in the particular state $|i\rangle$ in a trial with polarization $\epsilon$.

To find matrix elements $M^d_{\epsilon i}$, we use the stochastic Markov chain approach. In particular, we perform simulation of the 8-level $^{138}$Ba$^+$ system and introduce the $9^{\rm th}$ state to keep track of the average number of scattered 493~nm photons~\cite{Crocker2019}. Here, we assume that the intensities are the same for all the polarizations of 493~nm beams, and that 650~nm beams for each polarization also have the same intensities. Under these assumptions, the resulting matrix  $M^d$ acquires the following form:
\begin{equation}
  M^d = 
  \begin{pmatrix} 
   6.6 & 5.4 & 0 & 0  \\
   0 & 0 & 5.4 & 6.6  \\
   0 & 6 & 6 & 0 \\
   13.3 & 12.6 & 11.4 & 0 \\
   0 & 11.4 & 12.6 & 13.3 \\
   \end{pmatrix}.
\end{equation}

Additionally, we impose two constrains. (i) the unitarity condition: $\sum^3_{i=0}d_i = 1$ \; and (ii) populations $d_i\in [0;1],\;\; i=0,\dots,3 $.
As a result, we construct an over-constrained problem with four unknown populations $d_i$, five equations, four bounds, and one constraint. 

Two different ways of solving this problem are discussed below, with the first one treating the efficiency $E_d$ as an additional unknown, while the second one taking it as an independently determined constant.

In the first method of solution, in addition to $\{d_i\}$ and $E_d$, we introduce another unknown variable $C_b$ -- a static background count rate. That results in the total of six unknown variables and six equations: five from different polarization configurations as in Eq.~(\ref{eq:5polar}) and the unitarity condition. This method, however, does not account for the bounds $d_i\in [0;1]$. The matrix form of the equation becomes:
\begin{equation}\label{eq:5polarandCb}
   \begin{pmatrix} 
   n_+ \\
     n_- \\
     n_{\pi}\\
     n_{+\pi}\\
     n_{-\pi}\\
   \end{pmatrix}
  =E_d\begin{pmatrix} 
   6.6 & 5.4 & 0 & 0 & 1  \\
   0 & 0 & 5.4 & 6.6 & 1 \\
   0 & 6 & 6 & 0 & 1\\
   13.3 & 12.6 & 11.4 & 0 & 1 \\
   0 & 11.4 & 12.6 & 13.3 & 1\\
   \end{pmatrix} 
   \begin{pmatrix} 
   d_0 \\
     d_1 \\
     d_2\\
     d_3\\
     C_b\\
   \end{pmatrix}.
\end{equation}

We solved this system directly in Ref.~\cite{Crocker2019}. Since we did not impose the requirements $d_i\in [0;1],\; i=0,\dots,3 $, some of the populations that we obtained that way fluctuated outside of this physical region due to the background noise and polarization imperfections. 

The second method takes all the necessary conditions, including the bounds, into account, making the system over-constrained. We solve it using a constrained linear least-squares solver. The data shown in the main text is obtained with this method of analysis, but the two methods give very closely matching results. The corresponding error bars are obtained via the Hessian calculated at the solution point.

%

\end{document}